\documentclass[aps,pra,
twocolumn,
amsmath,amssymb,
]{revtex4}
\usepackage{graphicx}% Include figure files
\usepackage{dcolumn}% Align table columns on decimal point
\usepackage{etoolbox}

\usepackage{amsfonts,bm}
\usepackage{graphicx,epstopdf}  %epsfig,
\usepackage{color}
\usepackage{extarrows}
\usepackage{empheq}
\usepackage{seqsplit}
\usepackage{enumerate}

\allowdisplaybreaks[1]

\renewcommand{\d}{{\rm d}}

\newcommand{\w}{\omega}
\newcommand{\wti}{\widetilde}

\newcommand{\fit}{{\rm fit}}

\newcommand{\dg}{\dagger}
\newcommand{\la}{\langle}
\newcommand{\ra}{\rangle}

\newcommand{\B}{\mbox{\tiny B}}

\newcommand{\BEFD}{\mbox{\tiny F/B}}

\newcommand{\tB}{\mbox{\tiny B}}

\newcommand{\tSB}{\mbox{\tiny SB}}

\newcommand{\tS}{\mbox{\tiny S}}

\newcommand{\T}{\mbox{\tiny T}}

\newcommand{\be}{\begin{equation}}
\newcommand{\ee}{\end{equation}}
\newcommand{\bsube}{\begin{subequations}}
\newcommand{\esube}{\end{subequations}}
\newcommand{\Eq}[1]{Eq.\,(\ref{#1})}

\newcommand{\Fig}[1]{Fig.\,\ref{#1}}

\usepackage{tikz}
\usetikzlibrary{positioning, shapes.geometric}
\newcommand{\RN}[1]{%
  \textup{\uppercase\expandafter{\romannumeral#1}}%
}

\begin{document}

\title{
	Universal Prony fitting decomposition for optimized hierarchical quantum master equations
}
%%%
\author{Zi-Hao Chen}
%\author{Yu Su}
\author{Yao Wang}
\author{Xiao Zheng}
\author{Rui-Xue Xu}
\author{YiJing Yan}
\email{yanyj@ustc.edu.cn}
\affiliation{Department of Chemical Physics, University of Science and Technology of China, Hefei, Anhui 230026, China}

\date{\today}

\begin{abstract}
	In this work, we propose the Prony fitting decomposition (PFD) as an accurate and efficient exponential series method, applicable to arbitrary interacting bath  correlation functions.
	The resulting hierarchical equations of motion (HEOM) formalism is greatly optimized, especially in extremely low temperature regimes that would be inaccessible with other methods.
	For demonstration, we calibrate the present PFD against the celebrated Pad\'e spectrum decomposition method, followed by converged HEOM evaluations on the single--impurity Anderson  model system.
\end{abstract}
\maketitle

Dissipation is ubiquitous in almost all realms of modern physics.
Various quantum dissipation theories (QDTs) are exploited to deal with the reduced system dynamics.
Almost all QDTs are based on the Gaussian environment description, whose influence can be completely dictated by the interacting bath correlation function, $C_{\B}(t)$.
Theoretical methods, such as hierarchical equations of motion (HEOM), \cite{Tan906676, Tan06082001, Yan04216, Xu05041103, Xu07031107, Jin08234703, Ye16608, Yan16110309}quantum Monte Carlo \cite{Ema11349} and numerical renormalization group, \cite{Oso13245135} are heavily based on the efficiency of exponential series expansion,
\be \label{decomp}
C_{\B}(t) \!= \!\frac{1}{\pi}\!\int_{-\infty}^{\infty}\!\!\d \w\, e^{\pm i\w t} J(\w) f^{\BEFD}(\w) \simeq \sum_{k=1}^{K} \eta_k e^{-\gamma_k t}.
\ee
% Here, $J(\w)$ is the spectrum density of interacting bath and $f^{\BEFD}(\w)$ is the Fermi or Bose function. 
% 
% We denote $\beta=1/(k_BT)$ with $k_B$ being the Boltzmann constant and $T$ the temperature. 
% 
% Throughout the paper we set $\hbar=1$.
% 
The first identity is the fluctuation--dissipation theorem, involving $J(\w)$, the interacting bath spectrum density, and $f^{\BEFD}(\w)$, the Fermi/Bose function. The second identity arises from certain sum--over--poles (SOP) scheme, followed by Cauchy’s contour integration. In convention, $f^{\BEFD}(\w)$ is the quantity of the SOP decomposition, since $J(\w)$ is often given with models.
By far the Pad\'{e} spectrum decomposition (PSD) is the golden standard scheme, \cite{Hu10101106,Hu11244106} except for the extremely low--temperature regime, due to the underlying discontinuity that leads to the accuracy length of effective $f^{\BEFD}(\w)$ shrinking intensely. Alternative methods include the Fano spectrum decomposition, \cite{Cui19024110,Zha20064107}
discrete Fourier series \cite{Zho08034106}
and extended orthogonal polynomials expansions. \cite{Liu14134106,Tan15224112,Nak18012109,
	Ike20204101,Lam193721}

In this work, we propose the Prony fitting decomposition (PFD) scheme, directly for the second identity of \Eq{decomp}.
As known, optimized Prony fitting method \cite{Bey0517} has been exploited in other fields of physics.
Note that $C_{\B}(t) \equiv C^{(r)}_{\B}(t) + i C^{(i)}_{\B}(t)$, with the temperature dependence appearing only in one of the parts. It is the imaginary part for the fermionic case, while the bosonic case is the opposite.
Naturally, the numerical fit is to be performed against the exact $C_{\B}(t)$ satisfing the first identity of \Eq{decomp}, the fluctuation--dissipation theorem.

\begin{figure}
	\centering
	\includegraphics[width=0.48\textwidth]{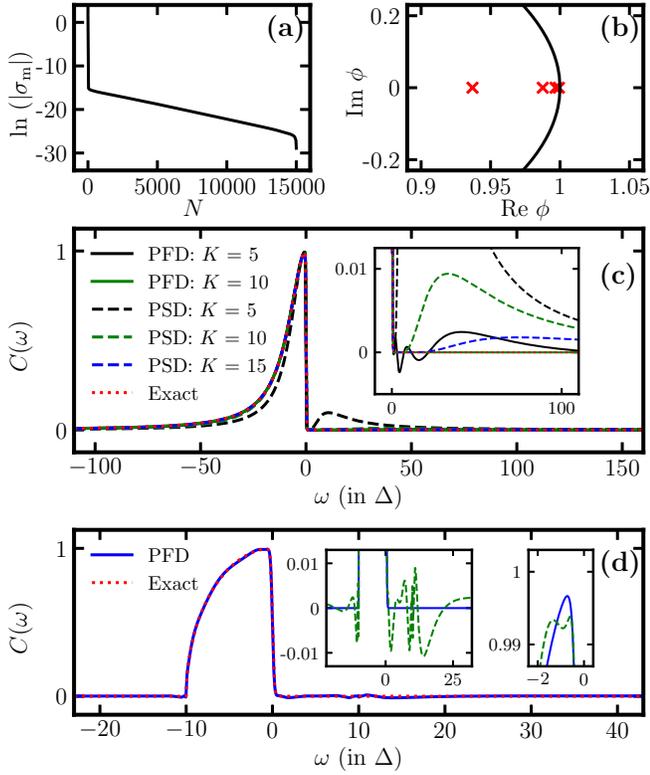}
	\caption{An illustrative example of PFD scheme. In panels (a), (b) and (c), the Lorentzian spectral density is adopted  with $W = 10\Delta$ and $\beta \Delta = 10$ [cf.\,\Eq{lor_def}]. In (d), the semicircle $J(\w)$ is adopted [cf.\,\Eq{semi}].
		In panels (a) and (b) are the illustrations of step (3) and (4) of the PFD scheme, respectively, 
		with $K_r = 1$  and $K_i = 4$.
		In panels (c) and (d), the fitting results are shown explicitly with the spectrum $C(\w)$ [cf.\,\Eq{c_w_def}].}
	\label{fig1}
\end{figure}

We will see that PFD is superb, covering extremely low--temperature regimes and also rather arbitrary spectral densities.
The resulting PFD--HEOM shows a greatly enhanced range of applications.
The proposed PFD, exemplified with the fermionic case detailed in the caption of \Fig{fig1}, targets at optimizing $K_{i}$ in
\be \label{decomp2}
C^{(i)}_{\B}(t) \equiv {\mathrm{Im}}\, C_{\B}(t) = \sum_{k=1}^{K_{i}} \zeta_k e^{- \lambda_k t}.
\ee
The total $K = K_{r} + K_{i}$, where $K_{r}$ is via the real part fitting.
Its protocol goes as follows.

\noindent (1) Sample the exact and temperature dependent $C^{(i)}_{\B}(t)$ to obtain
\begin{equation}
	\phi_j \equiv C_{\B}^{(i)}(j \d t);\ \ j = 0,\cdots,2N .
\end{equation}
Here, $\d t \equiv t_{\rm c}/(2N)$ with $t_{\rm c}$ being the cutoff time of $C^{(i)}_{\B}(t)$. For demonstrations, we set $N=15000$ and $t_{\rm c} = 80\Delta^{-1}$, where $\Delta$ stands for the characteristic width of $J(\w)$; see the caption of \Fig{fig1} for the details.

\noindent (2) Construct the $(N+1) \times (N+1)$ Hankel matrix,
\begingroup
\renewcommand*{\arraystretch}{1.5}
\begin{align}
	\label{Hankel_matrix}
	\bf{H} =
	\begin{bmatrix}
		\phi_0 & \phi_1     & \cdots & \phi_N     \\
		\phi_1 & \phi_2     & \cdots & \phi_{N+1} \\
		\cdots & \cdots     & \cdots & \cdots     \\
		\phi_N & \phi_{N+1} & \cdots & \phi_{2N}
	\end{bmatrix}.
\end{align}
\endgroup

\noindent (3) Decompose $\bf H$ with the Takagi’s factorization,
\begin{equation}
	{\bf H u}_m = \sigma_{m} {\bf u}^{*}_m;\ \ m=0,\cdots, N.
\end{equation}
Here, ${\bf u}_m^{*}$ is complex conjugate of ${\bf u}_m \equiv \{u_{0m}, \cdots\!, u_{Nm} \}$.
The obtained $\sigma_m$ and ${\bf u}_m$ are called c-eigenvalues and c-eigenvectors in literature.
We order $\{\sigma_m\}$ according to the descending $|\sigma_m|$--values and $\{{\bf u}_m\}$ just follows.
In most cases, $\{|\sigma_m|\}$ descend rapidly; see \Fig{fig1}(a). %the Prony method we use depends on the fast decay of those c-eigenvalues\cite{Bey0517};

\noindent (4) Upon the above setup, we select ${\bf u}_{m=K_{i}}$ to define
\be
f(z) \equiv \sum_{n=0}^{N} u_{nK_{i}} z^n.
\ee
Then obtain the $N + 1$ roots, $w_0, \cdots\!, w_{N}$, of this polynomial. As known \cite{Bey0517}, $K_{i}$ roots are of $|w_{k}| < 1$, whereas others are setting in the unit circle; see red cross marks in \Fig{fig1}(b).
Now we obtain the exponents $\{\lambda_1, \cdots, \lambda_{K_{i}}\}$  in \Eq{decomp2} as
\begin{align} \label{ploy}
	\lambda_k & = - \frac{2N}{t_c} [\ln | w_k| + i \arg(w_k)],
\end{align}
where $\arg(w_k) \in (-\pi, \pi]$.

\noindent (5) Obtain $\{\zeta_1, \cdots, \zeta_K\}$ in \Eq{decomp2} by the least--squares fitting the following $2N+1$ equations,
\begin{align} \label{equation_obtain_w}
	\phi_j = \sum_{k=1}^{K_{i}} \zeta_k w_k^j;\ \ j = 0, \cdots ,2N.
\end{align}
Involved are those $K_{i}$ roots of $|w_k| < 1$; see the remarks after \Eq{ploy}.

\noindent (6) Treat $C_{\B}^{(r)}(t)$, following the same procedure above.
Resulting in total $K = K_{r} + K_{i}$ terms in \Eq{decomp}.
This finalizes the PFD steps.

\begin{figure}
	\centering
	\includegraphics[width=0.48\textwidth]{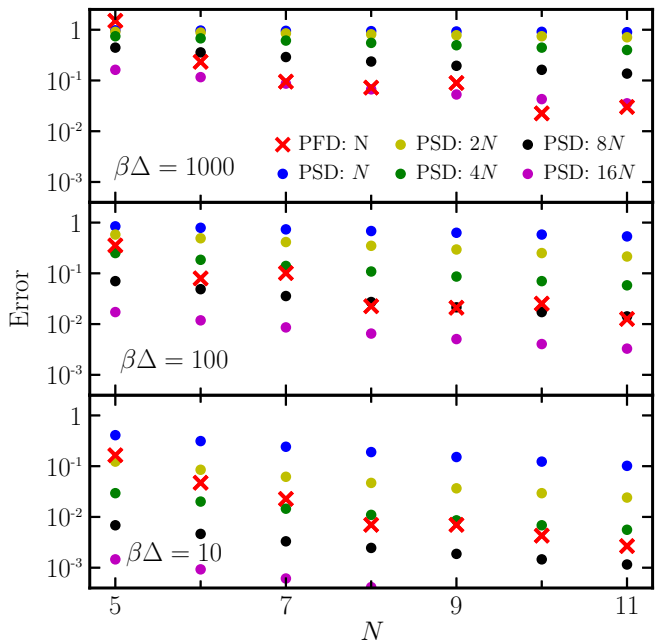}
	\caption{Performances of the PFD compared with PSD at different temperatures: $\beta \Delta = 1000$, $100$ and $10$. We adopt the Lorentzian form of spectral density  with $W = 10\Delta$.}
	\label{fig2}
\end{figure}

Some worthy remarks are as follows: 
(i) The exponents, \{$\lambda_{k}$\} in \Eq{decomp2} with \Eq{ploy}, are either real or in complex conjugate pairs.
This property is required by the underlying time--reversal symmetry in \Eq{decomp}; 
(ii) Generally speaking, both $C^{(r)}_{\B}(t)$ and $C^{(i)}_{\B}(t)$ require PFD, except for some special cases. Demonstrated in \Fig{fig1}(c) and (d) are the Lorentzian and semicircle spectral densities. In the former case, the $C^{(r)}_{\B}(t)$ can be obtained as a single exponential term analytically, and needs no fitting treatment. This simply leads to $K_r = 1$. While in the later case, both $C^{(r)}_{\B}(t)$ and $C^{(i)}_{\B}(t)$ need fitting;
(iii) The PFD results for Lorentzian spectral density,
\begin{equation}\label{lor_def}
	J(\w) = \frac{\Delta W^2}{\w^2 + W^2},
\end{equation}
compared with the PSD schemes, are illustrated in \Fig{fig1}; see black the solid and dash lines in panel (c).
Exhibited in \Fig{fig1}(c) are also other different choices of $K$. They are explicitly shown with the spectrum,
\be \label{c_w_def}
C(\w)=\frac{1}{2}\int_{-\infty}^{\infty}\!\!{\rm d}t\,e^{-i\w t} C_{\B}(t).
\ee
It can be found that errors primarily occur near the zero frequency.
We will compare the performances of PFD and PSD quantitatively below in \Fig{fig2}.

\begin{figure}[t!]
	\centering
	\includegraphics[width=0.48\textwidth]{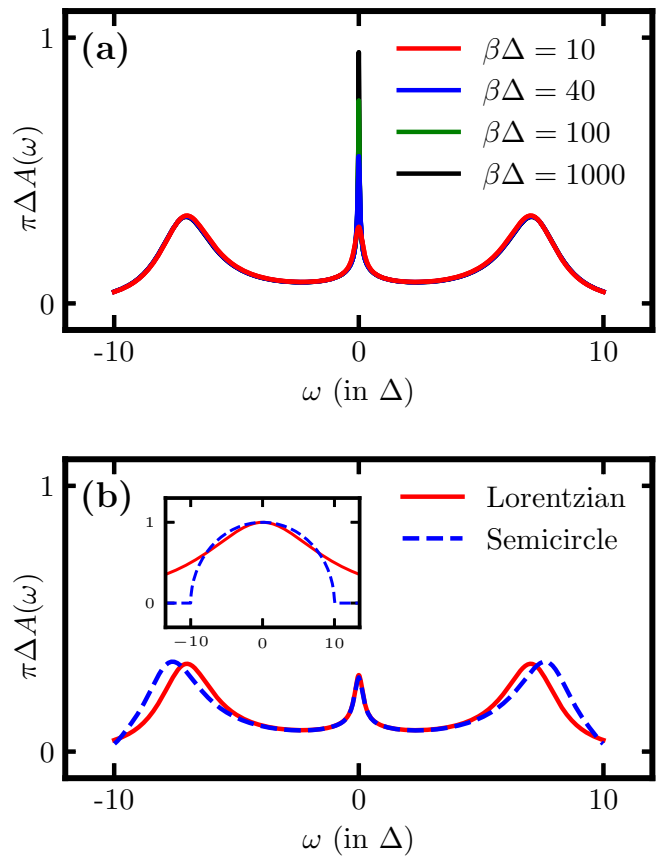}
	\caption{The HEOM simulation results of $\Delta A(\w)$ with different types of spectral densities.
		In panel (a), we apply the Lorentzian spectral density with $W = 10 \Delta$. Temperatures are $\beta \Delta = 10$, $40$, $100$ and $1000$. Other parameters are: $U = 12 \Delta$  and $\epsilon = - U /2$.
		In panel (b) we compare the  result of $A(\w)$ with the Lorentzian versus semicircle spectral densities (cf.\,the inserted subfigure) in the case of $\beta \Delta = 10$.}
	\label{fig3}
\end{figure}

As an example of non-analytical spectral density cases, we also apply the PFD scheme to semicircle spectral density [cf.\,the inserted subfigure of \Fig{fig3}(b)],
\be \label{semi}
J(\w)=
\begin{cases}
	\Delta \sqrt{1- (\w/W)^2}, & -W \leq \w \leq W,      \\
	0,                         & \w<-W\ \text{or}\ \w>W.
\end{cases}
\ee
The fitting result are shown in \Fig{fig1}(d). Notice that the PSD method is absent due to its failure in this case of spectral density.
We fit the real part with $K_r = 7$ and the imaginary part with $K_i = 8$, respectively.
%
% We observe four extremely small pre-exponential factors, and one may use $K=K_r+K_i-4=11$ terms in \Eq{decomp} to further reduce the computational costs.

In \Fig{fig2}, we calibrate the relationships between the accuracy and $K$ for Lorentzian spectral density at different temperatures.
The fitting errors are measured by
\begin{equation}
	\mathrm{Error} = \frac{\int_{-\infty}^{\infty}\, \d \w \left|C_{\tB}^{\fit}(\w) - C_{\tB}(\w) \right|}{\int_{-\infty}^{\infty}\, \d \w \left|C_{\tB}(\w)\right|}.
\end{equation}
We observe that at the same accuracy level, PFD requires $1/16$, $1/8$ and $1/4$ summation terms of that required by PSD when $\beta \Delta = 1000$, $100$ and $10$, respectively.
This is exactly what we need for overcoming the low--temperature curse encountered in HEOM simulations. The PFD scheme remarkably reduces the computational and memory costs.
In the fermionic HEOM evaluations, the number of involved auxiliary density operators is
\begin{equation}\label{number}
	\mathcal{N} = \sum_{l = 1}^{{L}}\frac{\wti K!}{l!(\wti K-l)!}
\end{equation}
where $L$ is the tier of hierarchy truncation and the total number of involved exponential terms is $\wti K=2\times N_{\alpha} \times N_{u}\times K$, with $N_{\alpha}$ and $N_{u}$ being the numbers of bath reservoirs and system orbitals, respectively. The factor of $2$ accounts for particles plus holes.
To be concrete, when the $L$ is $6$, PFD requires only $10^{-7}$, $10^{-5}$, $10^{-4}$ computational and memory costs of that required by PSD when $\beta \Delta = 1000$, $100$ and $10$ [cf.\,\Eq{number}].

As a numerical demonstration, we exploit the PFD scheme in the HEOM simulations of SIAM. \cite{Lia02725,Far20256805,Moc21186804,Kur216004,Fer20738}
SIAM is frequently exploited to study the
the Kondo resonance, which is massively investigated since it manifests strong electronic correlations at very low temperatures.
Its  Hamiltonian reads
\begin{equation} \label{aim_def}
	H_{\T} = H_{\tS} + H_{\tSB} + h_{\tB},
\end{equation}
where the system is
\be
H_{\tS}  = \epsilon (\hat n_{\uparrow} + \hat n_{\downarrow}) + U \hat n_{\uparrow} \hat n_{\downarrow}
\ee
with
$\hat n_{s}=\hat a_{s}^{\dg}\hat a_{s}$,  the system--environment interaction reads
\begin{equation}
	H_{\tSB} =\sum_k \sum_{ s={\uparrow,\downarrow}} t_{ k s} (\hat a_{ s}^{\dagger} \hat d_{ k s} + {\rm H.c.}),
\end{equation}
and the environment,
$
	h_{\tB} = \sum_{ k s} \epsilon_{ k s} \hat d_{ k s}^{\dagger} \hat d_{ k s},
$
is composed of free elections.
In simulation, we select SIAM parameters to be $U = 12 \Delta$ and $\epsilon = -U/2$, and the Kondo temperate is around $\beta \Delta = 40$.

The simulation results are exhibited with impurity spectral density
\begin{equation}
	A_{s}(\w) = \frac{1}{2 \pi} \int_{-\infty}^{\infty}\!\!\d t\, e^{i\w t} \la\{ \hat a_{s} (t), \hat a_{s}^{\dagger} (0)\} \ra,
\end{equation}
where $\hat a_{s}$ ($\hat a_{s}^{\dagger}$) is the creation (annihilation) operator of the  electron with spin $s$.
The HEOM simulation results of $\Delta A(\w)$ with Lorentzian spectral densities are shown in \Fig{fig3}(a) at different temperatures: $\beta \Delta = 10$, $40$, $100$ and $1000$ are shown in \Fig{fig3}.
In these simulations, the number of exponential terms, $K$ in \Eq{decomp}, is $5$, $6$, $8$ and $10$ respectively, and
we set the truncation tier to be $L = 6$.
As expected, the Kondo peak $\pi \Delta A(0)$ increase to $1$ as the temperature decrease to $0$.
This behavior agrees with the Friedel sum rule \cite{Lan66516} at zero temperature, $A(0) = \sin^2(\pi \bar n) / (\pi \Delta)$, where $\bar n$ is the average electron occupation number.
It is also observed that the Hubbard peaks occurs at $\w = \pm U/2$, and the these peak are almost not affected by temperature.
In \Fig{fig3}(b), we compare the simulation results of $A(\w)$ with the Lorentzian versus semicircle spectral densities.
As shown in the figure, the difference appears apparently near the Hubbard peaks.

In summary, we propose the PFD scheme to accurately decompose the environment correlation functions into exponential sums.
This scheme significantly improves the efficiency and applicability of HEOM.
It not only helps reduce the numerical costs at extremely low temperatures, but also enables HEOM to deal with analytical spectral densities.
We exhibit and calibrate the PFD scheme in fermionic scenario and take the SIAM as an example.
It is anticipated that the PSD scheme will greatly benefit the HEOM simulation at cryogenic temperatures.

Support from
the Ministry of Science and Technology of China (Grant No.\ 2017YFA0204904 and 2021YFA1200103) and the National Natural Science Foundation of China (Grant Nos.\ 22103073 and 22173088) is gratefully acknowledged. Y. Wang and Z. H. Chen thank also the
partial support from GHfund B (20210702).
We are indebted to Yu Su for valuable discussions.
%
%The data that support the findings of this study are available from the corresponding author upon reasonable request.

%\bibliographystyle{./aiptit.bst}
%\bibliography{./bibrefs.bib}

\end{document}